# STEM image analysis based on deep learning: identification of vacancy defects and polymorphs of MoS$_2$

*Kihyun Lee[1,‡], Jinsub Park[1,‡], Soyeon Choi[1,‡], Yangjin Lee[1,2], Sol Lee[1,2], Joowon Jung[1], Jong-Young Lee[3], Farman Ullah[4], Zeeshan Tahir[4], Yong Soo Kim[4], Gwan-Hyoung Lee[3,5,6,7], and Kwanpyo Kim[1,2,*]*

[1]Department of Physics, Yonsei University, Seoul 03722, Korea.

[2]Center for Nanomedicine, Institute for Basic Science (IBS), Seoul 03722, Korea.

[3]Department of Material Science and Engineering, Seoul National University, Seoul, 08826, Korea.

[4]Department of Physics and Energy Harvest Storage Research Center, University of Ulsan, Ulsan 44610, Korea.

[5]Research Institute of Advanced Materials, Seoul National University, Seoul, 08826, Korea.

[6]Institute of Engineering Research, Seoul National University, Seoul, 08826, Korea.

[7]Institute of Applied Physics, Seoul National University, Seoul, 08826, Korea.

**ABSTRACT**

Scanning transmission electron microscopy (STEM) is an indispensable tool for atomic-resolution structural analysis for a wide range of materials. The conventional analysis of STEM images is an extensive hands-on process, which limits efficient handling of high-throughput data. Here we apply a fully convolutional network (FCN) for identification of important structural features of two-dimensional crystals. ResUNet, a type of FCN, is utilized in identifying sulfur vacancies and polymorph types of $MoS_2$ from atomic resolution STEM images. Efficient models are achieved based on training with simulated images in the presence of different levels of noise, aberrations, and carbon contamination. The accuracy of the FCN models toward extensive experimental STEM images is comparable to that of careful hands-on analysis. Our work provides a guideline on best practices to train a deep learning model for STEM image analysis and demonstrates FCN's application for efficient processing of a large volume of STEM data.

The identification and control of structural variety are key for understanding the properties of various nanomaterials and their applications. For two-dimensional (2D) crystals, the presence of point defects or grain boundaries, and the existence of different polymorphic configurations for a given chemical composition ratio strongly influence the electrical, chemical, and mechanical properties[1-3]. Recently, transmission electron microscopy (TEM) has been employed as an invaluable tool for atomic-resolution structural analysis and has indeed identified important structural features of 2D crystals[1]. For example, vacancies[4-7], topological defects[8], grain boundaries[9-13], different stacking configurations[14, 15], and stacking boundaries[16, 17] in graphene and transition metal dichalcogenides (TMDCs) have been extensively investigated. The visualization of structural variety in 2D crystals along with direct structure-property correlation has been developed by TEM and scanning transmission electron microscopy (STEM) imaging, which advanced our fundamental understanding of 2D materials and reliable adaptation toward electronic and other applications. However, conventional TEM or STEM image analysis requires extensive human input and is often time-consuming, which makes it inadequate for high-throughput data analysis. Considering the dramatically increasing quality and quantity of TEM data[18-20], the development of an image analysis process compatible with high-throughput data is of great utility.

Recent advancements in deep learning, especially the advent of convolutional neural networks (CNNs), holds great promise for feature recognition and analysis from TEM data[18-22]. Indeed, CNN-based image analysis has recently been applied to identification of various structural features in TEM and STEM image analysis[21, 23-32]. CNNs have been successfully adapted for identification of point defects and their evolution under e-beam irradiation, denoising of TEM/STEM images, sub-angstrom reconstruction around point defects, and structural phase evolution[21, 23-32]. Although these previous studies have clearly demonstrated the promise of CNNs, there are still many challenges to overcome. One of the most pressing challenges is the major bottleneck in the range of TEM/STEM image quality that CNNs can reliably handle. In practice, images with a low signal-to-noise ratio and with significant carbon contamination are discarded and only carefully selected images are currently processed with CNNs. Therefore, building a CNN model that is robust to intrinsic and extrinsic flaws in TEM data would be extremely beneficial.

Here we apply a fully convolutional network (FCN) for the identification of important structural features in atomic-resolution STEM images of 2D crystals. We create an FCN to process $MoS_2$, the archetype of TMDCs, to demonstrate its utility for the identification of sulfur vacancies and different polymorphic configurations. ResUNet[24], a type of FCN, was trained primarily with simulated STEM images containing different levels of noise, aberrations, and carbon contamination. The model prediction for sulfur vacancies was tested with experimental STEM images of as-synthesized and intentionally defect-created $MoS_2$ samples, which confirmed that the model accuracy for vacancy assignment is comparable to labor-intensive analysis based on the local intensity at atomic sites. Another FCN is trained for polymorph identification and successfully identified various stacking polymorphs from experimental STEM images of $MoS_2$. In regions with serious carbon contamination, the trend of polymorph model prediction follows an expectation based on local intensity modulation of atomic sites. Our work extends the utility of FCNs for analysis of a large volume of STEM images, including images with carbon residue.

As benchmark tasks, we perform identification of 1) sulfur vacancy defects and 2) polymorph types from atomic resolution STEM images of $MoS_2$ as shown in Figure 1. These structural features are important for determining the various properties of TMDCs, including $MoS_2$. For example, the sulfur vacancy is a predominant point defect in $MoS_2$ and has a strong influence on the material's doping level, charge transport behavior, and chemical reactivity[7, 33-36]. Previous deep-learning-based analysis on chalcogen vacancies in TMDCs was mostly on Se or Te[24, 31, 32] and S vacancies were rarely analyzed. The lower Z-contrast of S atoms in STEM images compared to Se or Te atoms may pose some challenges in precise vacancy assignment from FCN. The intra-layer structure and interlayer stacking configurations of TMDCs also determine the electronic bandstructures of materials[3, 37-42]. Therefore, we focus on identification of single- (named $VS_1$) or double-sulfur vacancies (named $VS_2$) as well as the different types of $MoS_2$ polymorphs (monolayer/bilayer/trilayer 2H-stacking, bilayer/trilayer 3R-stacking configurations, and 1T-stacking) as shown in Figure 1. To our knowledge, the identification of stacking polymorphs in TMDCs based on deep learning analysis is yet to be reported.

We used a ResUNet architecture (Figure 2c) for both vacancy identification and polymorph detection models. In training FCN models, a sufficiently large set of data with enough diversity and proper quality is crucial[19, 20]. As shown in Figure 2a,b and Supporting Figure S1, we mainly trained our ResUNet models with simulated STEM images (256×256 pixels) generated by Computem[43]. Some experimental images were also incorporated into the training and validation sets for the polymorph model. The simulated STEM images were post-processed to add intrinsic and extrinsic image distortions to reflect realistic imaging and sample conditions. We paid particular attention to the effects of shot noise and random vibration noise (Supporting Figures S2 and S3), carbon contamination (Supporting Figure S4), and additional noise source (Supporting Figure S5) on the model's performance with simulated and experimental images. ResUNet models that are properly trained with simulated images successfully recognize the important structural features from simulated images as well as experimental data. More details on the training process can be found in Methods and Supporting Information.

There are some important features in the model training. Figure 3 shows the monitored training process (Figure 3a-d) and the trained model's predictions on simulated test images (Figure 3e-l) from the vacancy and polymorph models. The vacancy in a crystal is an imbalanced dataset, where the number of vacancies is much less than the number of non-vacancies. Although model training based on the F1 score is more desirable for imbalanced classification, we chose to train the vacancy model using the standard accuracy and loss metric due to realistic consideration of the overhead required to train the vacancy model with user-defined metrics on Keras. As shown in Figure 3a, there are three jumps within the first 100 epochs of training in accuracy plots, which was consistently observed in the training process. By comparing the predictions right before and after the jumps, we could visualize changes in model prediction patterns at these particular epochs. We found that, after the first jump, the model stopped random predictions of vacancy locations and identified all points as atoms (without vacancies), which is expected for model training with an imbalanced dataset based on accuracy. After the second jump, the model started producing meaningful predictions of vacancy locations. The training and validation accuracy with simulated STEM images reach over 0.98 and the loss functions decrease to the level of 0.01 under the training of 1,000 epochs. The inspection of model predictions also confirms that reliable identification of $VS_1$ and $VS_2$ defects are achieved as shown in Figure 3e-h.

We trained the polymorph model with simulated STEM images of 6 polymorphs (monolayer/bilayer/trilayer 2H-stacking, bilayer/trilayer 3R-stacking, and 1T-stacking configurations). Individual images used for training and validation were either single type or had two different polymorphs with a polymorph boundary therein. (Figure 3i-l). Without the issues from imbalanced data sets, the accuracy curves for the polymorph models increase smoothly without jumps as shown in Figure 3c. The training and validation accuracies reach over 0.95 and the loss functions decrease below 0.4 with 50-epoch training. The training was terminated for both vacancy and polymorph models when the validation loss did not show any further decrease as shown in Figure 3b and 3d. It should be noted that the somewhat large loss function values obtained for the polymorph model compared to the vacancy model are due to the use of a different loss function (categorical cross entropy loss) and some misidentification near the polymorphic boundaries in images. Nonetheless, the polymorph model reliably predicts the types of $MoS_2$ polymorphs as shown in Figure 3i-j.

The trained models were applied to experimental STEM images for identification of vacancy defects and the types of polymorphs. Figure 4 summarizes the results of vacancy model prediction and performance comparison with conventional hands-on analysis. Chemical vapor deposition was utilized to grow monolayer $MoS_2$ and the samples were transferred to TEM grids for STEM imaging[43, 44]. We compare two kinds of samples: as-synthesized $MoS_2$ and hydrogen plasma-treated $MoS_2$ with intentional vacancy formation (Figure 4a and 4f). Remote hydrogen plasma treatment has recently demonstrated its utility to introduce sulfur vacancy defects in $MoS_2$ under a controlled manner[43]. Figure 4b and 4g are exemplary input STEM images of as-synthesized $MoS_2$ and plasma-treated $MoS_2$, respectively. The locations of vacancies were predicted by the vacancy model as shown in Figure 4c and 4h, where the green dots represent single sulfur vacancies ($VS_1$) and blue dots represent double sulfur vacancies ($VS_2$). Model predictions show the overall increase of vacancy defects in the plasma-treated sample. The visual monitoring of vacancy model prediction can be confirmed in the zoomed-in STEM images (Figure 4d,e,i,j).

We carefully validate the performance of our FCN model with comparison to conventional analysis based on local intensity at atomic sites. High-angle annular dark-field (HAADF) STEM imaging shows the Z-contrast between different atomic sites[15, 45] and local sulfur vacancies can be identified as sites with lower intensity. For conventional analysis, we developed another FCN model to localize atoms from the images[25] (Supporting Note 2 and Supporting Figure S6) and extracted the intensity from the localized atomic sites. As shown in Figure 4k and 4m, the intensity histogram at localized atomic sites typically shows two strong peaks; a peak at the center position of 1.0 is from double occupancy of sulfur atoms and another peak at around 1.5 is from Mo atom sites. A small shoulder peak at a center of around 0.7 is attributed to $VS_1$ sites[43]. The fact that the shoulder peak increases in the plasma-treated samples is consistent with this interpretation. Rare events located below 0.5 can be assigned to the presence of $VS_2$. Some overlap between the $VS_1$ peak and pristine doubly-occupied S sites indicates that the identification of vacancies from experimental STEM images involves a level of uncertainty. Instrumental resolution set by STEM imaging conditions and extrinsic factors such as local carbon contamination may also have some influence on the observed intensity distribution. Due to the uncertainty regarding the assignment of ground-truth vacancy locations, we cannot calculate accuracy and error functions with a predetermined label image. On the other hand, the overall vacancy concentrations, which is robust to the observed intensity overlap, can be used to assess the model performance.

We compare the vacancy model prediction with the conventional hands-on analysis. To statistically analyze the FCN model prediction, 80 experimental images of as-synthesized $MoS_2$ samples (256×256 pixels) and 112 images of plasma-treated $MoS_2$ images were analyzed. Figure 4l and 4n show the accumulated model prediction shown in a manner similar to the intensity histograms. The segregated data set from model predictions (atomic sites, $VS_1$, and $VS_2$) in the spectrum line plots exhibits distributions consistent with conventional analysis, which reconfirms the reliable demonstration of the FCN vacancy model. The vacancy concentrations estimated by the FCN model and conventional analysis show the expected linear correlation as shown in Figures 4o and 4p. With the handling capability of large-throughput TEM data and comparable performance of defect concentration estimation, the vacancy FCN model can be a very powerful tool to monitor defect concentration and their heterogeneous distribution of defects in large-area data sets.

The FCN model can be applied to the identification of various stacking polymorphs from experimental STEM images. Figure 5 summarizes results from the polymorph model prediction with experimental STEM images of $MoS_2$. Various experimental stacking configurations were identified by conventional intensity-based analysis (Supporting Figure S7)[14]. We applied our FCN model to experimental STEM images (256×256 pixels) of polymorphs (2H monolayer: 416 images, 2H bilayer: 128 images, 3R bilayer: 364 images, 3R trilayer: 132 images). We note that $MoS_2$ samples with 1T or 2H trilayer structures were not available for our study. Some experimental images were seriously contaminated with local carbon accumulation but were included in the validation process. We found that the FCN polymorph model is quite robust to certain levels of carbon contamination and correctly identifies the true stacking configurations as shown in Figure 5a~c. We found that the accuracy of model prediction for 2H bilayer and 3R trilayer stacking configurations is quite high whereas 2H monolayer and 3R bilayer configurations exhibit somewhat lower accuracy. The visual inspection of the model prediction results showed that local areas with serious carbon contamination are often mislabeled (Figure 5c).

To understand the possible origin of misidentification from the polymorph model, we investigate the trend of model prediction in regions with different levels of carbon contamination (Figure 5d). Figure 5d shows a typical intensity histogram of each pixel in 2H monolayer $MoS_2$ STEM images after the application of Gaussian blurring. Therefore, the pixel intensity is positively correlated with local carbon contamination. The histogram often contains a long tail toward the high intensity region above 100, which can be assigned to regions with higher levels of carbon contamination. The plotted model prediction indicates that the performance of the model is quite high (above 90%) in the regions of pixel intensity below 100 but tends to mislabel the polymorph in the regions with more carbon contamination. With added background intensity from carbon contamination, the contaminated region can display a different local intensity pattern from its original configuration, so that it is easily misidentified as a different configuration. For example, the 2H monolayer configuration can exhibit local Mo:S intensity ratios similar to that of 3R bilayer and 3R trilayer configurations under heavier carbon contaminations as shown in Supporting Figure S8. This trend is consistent with polymorph model predictions. The results of model prediction with 3R bilayer images as a function of carbon contamination also confirm a similar trend as shown in Supporting Figure S9.

The accuracy of polymorph model prediction can be improved by limiting our analysis to regions without significant carbon contamination. We developed an independent carbon model based on FCN, which identifies the degree of local carbon contamination as shown in Figure 5e and Supporting Figure S10. In particular, we find that the polymorph model accuracy for 2H monolayer and 3R bilayer strongly depends on the degree of carbon contamination as shown in Supporting Figure S11. After the automated area selection with the aid of the carbon model, the polymorph model accuracy was significantly improved as shown in Figure 5f. The model accuracy is above 95 % for three types of polymorphs and 88 % for the 3R bilayer case.

In conclusion, we utilized ResUNet for reliable identification of sulfur vacancies and polymorph types in STEM images of $MoS_2$. The training of ResUNet models with simulation images in the presence of different levels of noise, aberrations, and carbon contamination enables the reliable identification of structural features in experimental STEM images. The performance of vacancy model is comparable to conventional time-consuming data analysis based on local intensity. The performance of polymorph models is also comparable to visual recognition based on local intensity. These results demonstrate that the developed models can be applied to a wide range of images with various levels of carbon contamination and other sources of noise. We believe that the current models can be extended to investigate similar structural features in various 2D crystals beyond $MoS_2$ and therefore provide a guideline on best practices to train a deep learning model for STEM image analysis.

**Methods**

**Model building and training:** We used a ResUNet architecture[24] with 5 residual blocks and 32 filters in the first convolutional layer, and a Keras Generator to load the training sets for both vacancy and polymorph detection. For training, we assigned a circular area with a radius of 0.95Å as sulfur vacancy labels. The vacancy-detection model used a kernel size of 7×7 pixels and 10,000 images for training and 5,000 images for validation. The generator loaded 1,536 images into the model every epoch and used a batch size of 32. Using a dropout rate of 0.5, an Adam optimizer with a learning rate of 0.01, and Dice loss function, the model was trained for 1,000 epochs. Monitoring the accuracy and loss curves indicated that training was usually complete after 150-200 epochs. Training beyond this point did not cause the model to overfit the training data and accuracies became saturated at the maximum values as shown in Figure 3 due to the large dropout rate of 0.5. The polymorph model used a kernel size of 7×7 pixels and 63,157 images for training and 17,319 images for validation. It used a dropout rate of 0.5, Adam optimizer with a learning rate of 0.0001, and categorical cross entropy loss function.

**STEM image simulation:** We used Computem[46] to simulate STEM images of $MoS_2$ (1024×1024 pixels). A range of different image sampling rates from 0.06 to 0.24 Å/pixel in increasing increments of 0.03 was utilized to obtain STEM images at different magnifications compatible with experimental observation. Simulated STEM images were post-processed to include various noise effects, including shot noise, random vibration noise, carbon contamination, and additional source of noise, and contrast adjustment. A detailed description of these processes can be found in the Supporting Information. Images were cropped to 256×256 pixels for model training and validation.

**Sample preparation and experimental STEM imaging**: $MoS_2$ samples were synthesized by chemical vapor deposition (CVD) processes. Detailed information on sample preparation can be found in previous publications[43, 44]. The remote hydrogen plasma-treatment for intentional introduction of sulfur vacancies was performed using a homebuilt indirect plasma system (L-Gen; Femto Science) embedded in a conventional CVD system. Detailed information about the treatment can be found in literature[43]. Experimental HAADF-STEM imaging was performed using an aberration-corrected TEM (ARM-200F; JEOL) operated at 80 kV with a probe current of 25 pA. The semi-convergence angle of the electron probe was 23 mrad and the collection semi-angles of the detector were ≈ 40-160 mrad. The dwell time was set to 16 µs and the original images were 1024×1024 pixels in size.

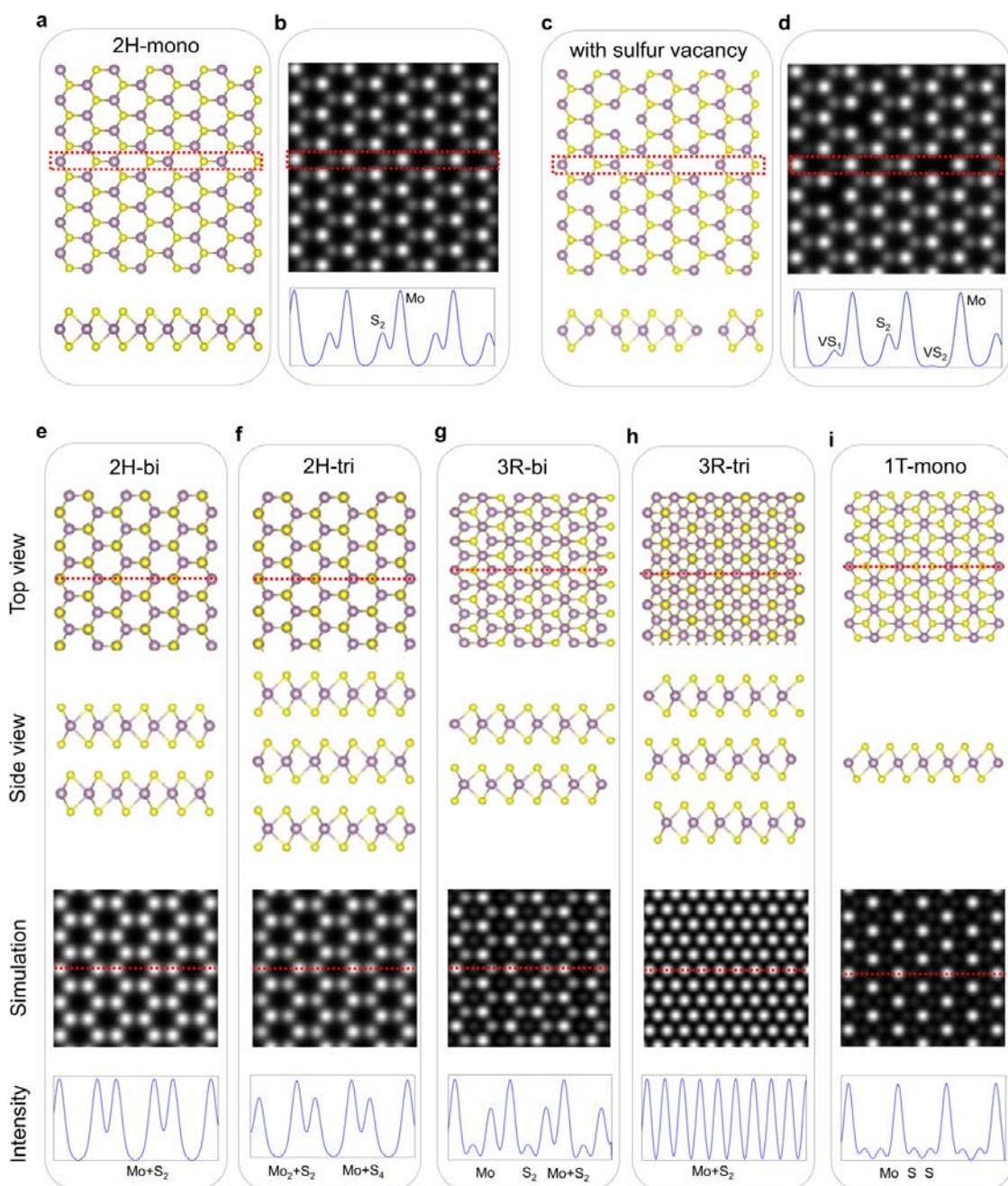

**Figure 1. MoS$_2$ polymorphs and their STEM image simulations.** (a) Top-view and side-view of monolayer 2H-MoS$_2$ atomic model. (b) Simulated STEM image of 2H-MoS$_2$ (top) and line intensity profile (bottom). (c) Top-view and side-view of monolayer 2H-MoS$_2$ atomic model with single sulfur vacancy (VS$_1$) and double sulfur vacancy (VS$_2$). (d) Simulated STEM image of 2H-MoS$_2$ with sulfur vacancies (top) and line intensity profile (bottom). Top-view, side-view, simulated STEM image, line intensity profile of MoS$_2$ from polymorphic configurations of (e) 2H bilayer, (f) 2H trilayer, (g) 3R bilayer, (h) 3R trilayer, (i) 1T monolayer.

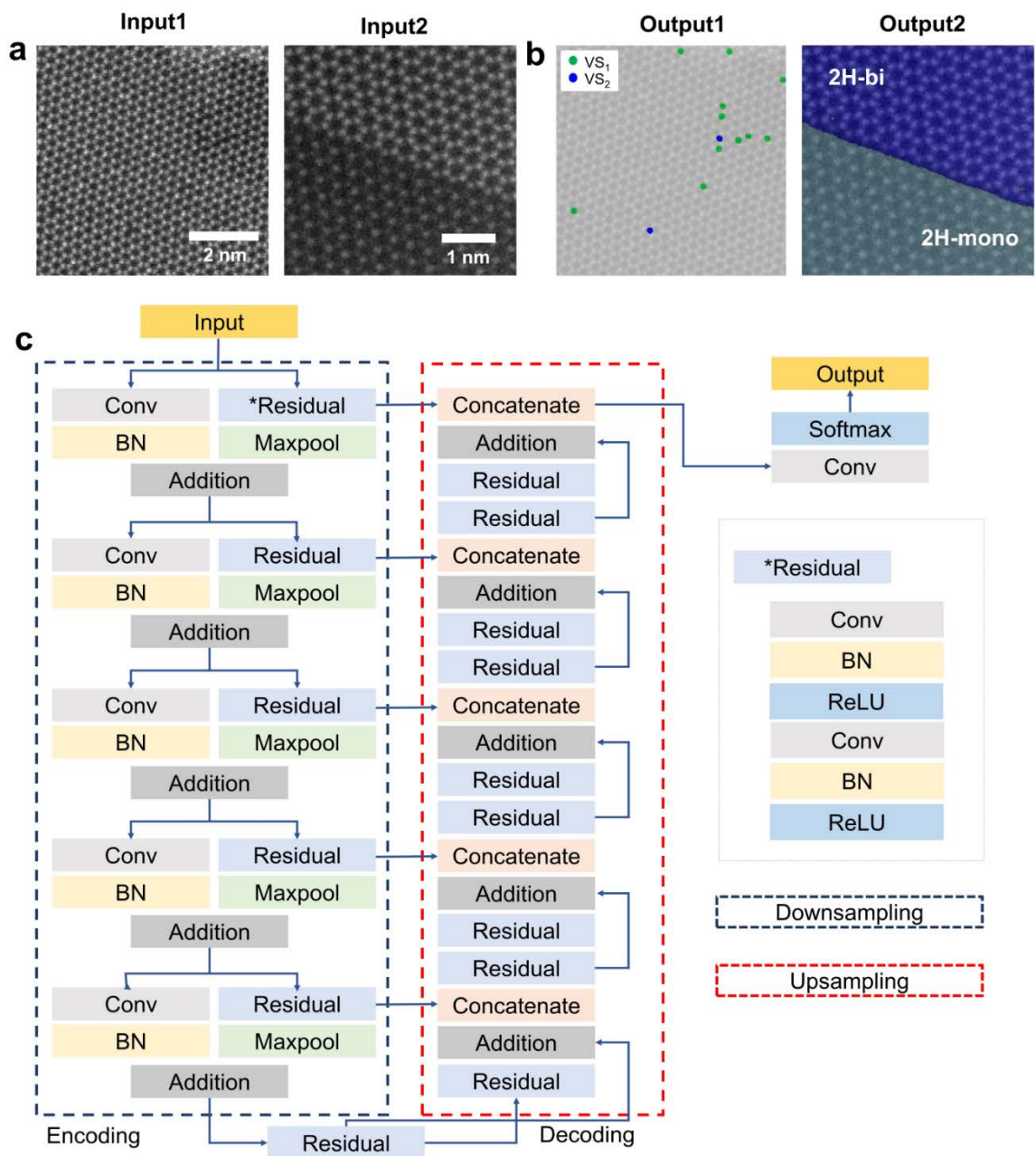

**Figure 2. Fully convolutional network (FCN) utilized for identification of point defects and polymorphs of MoS$_2$.** (a) Examples of input images (256×256 pixels) for point-defect identification (left) and polymorph identification (right). (b) Examples of output images (256×256 pixels). The different colors in the images note identified vacancies (left) or different regions of polymorphs (right). (c) Schematic of FCN used for identification of point defects and polymorphs. Conv, BN, Maxpool and ReLU indicate convolution, batch normalization, max pooling and rectified linear unit activation function, respectively.

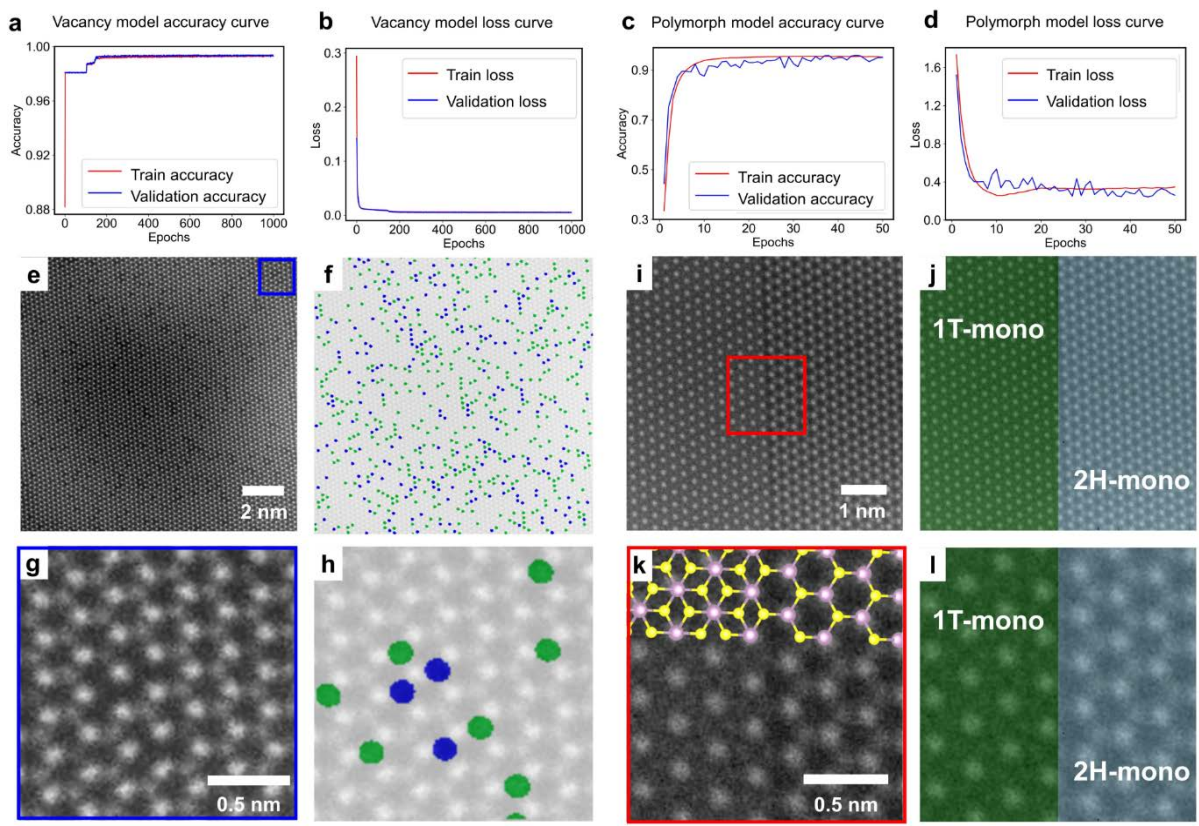

**Figure 3. Training and validation of FCN models with simulated STEM images.** (a) Training and validation accuracy curves for identification of vacancy defects. (b) Training and validation loss curves for vacancy model. (c) Training and validation accuracy curves for identification of polymorphs. (d) Training and validation loss curves for polymorph model. (e) Simulated STEM image of $MoS_2$ with scattered vacancy defects. The blue box is the field of view for the zoomed-in image of panel g. (f) Prediction results for vacancy defects ($VS_1$: blue, $VS_2$: green) from vacancy models. (g) Zoomed-in simulated image of $MoS_2$. (h) Zoomed-in prediction results for vacancies. (i) Simulated image of $MoS_2$ with two different polymorphs (2H monolayer and 2H bilayer). The red box is the field of view for the zoomed-in image of panel k. (j) Prediction from polymorph model. (k) Zoomed-in simulated image of $MoS_2$ with different polymorphs. (l) Zoomed-in prediction results from polymorph model.

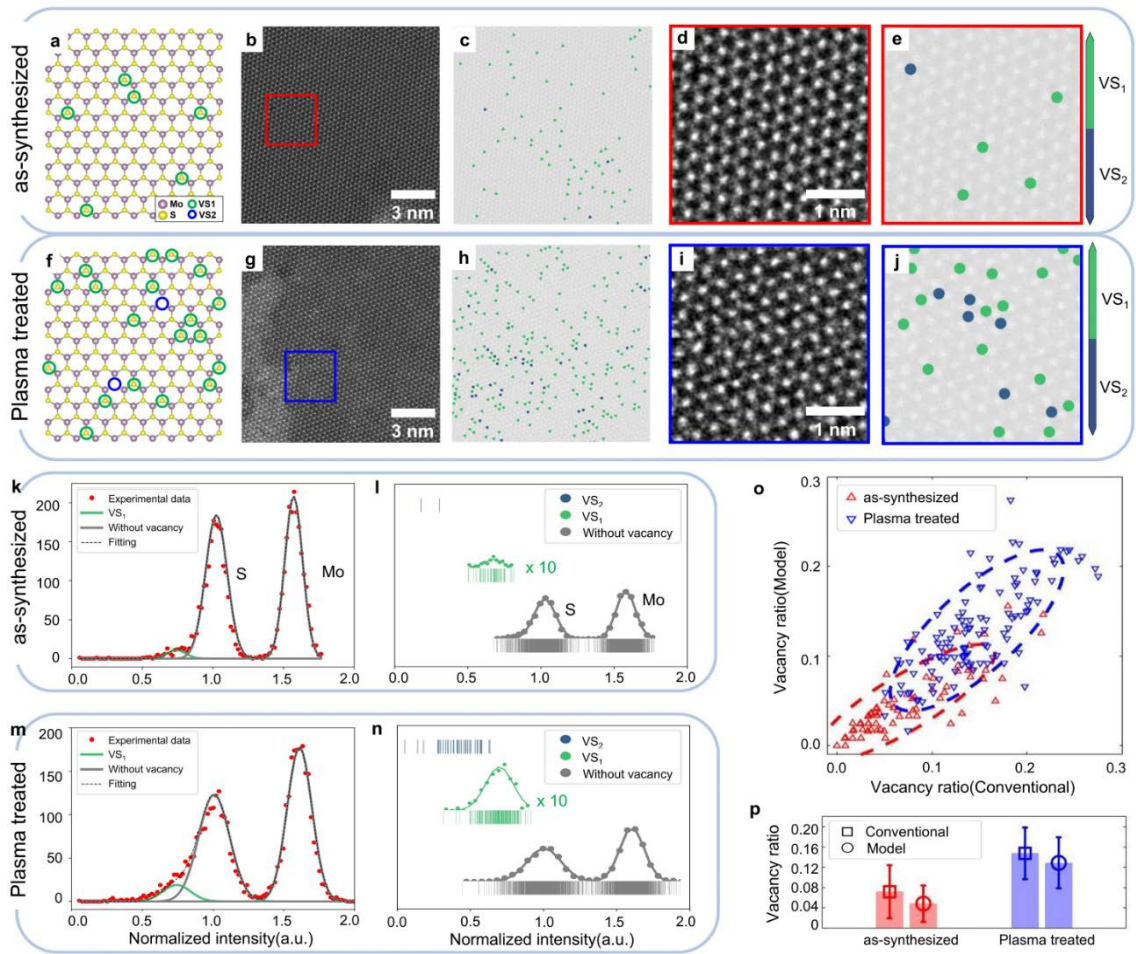

**Figure 4. Sulfur vacancy identification with experimental STEM images of MoS₂.** (a) Atomic model of as-synthesized MoS$_2$. (b) Experimental STEM image of as-synthesized MoS$_2$. The red box is the field of view for zoomed-in image of panel d. (c) Model prediction of sulfur vacancies. (d) Zoomed-in STEM image of as-synthesized MoS$_2$. (e) Zoomed-in model prediction of sulfur vacancies. (f) Atomic model of plasma-treated MoS$_2$. (g) Experimental STEM image of plasma-treated MoS$_2$. The blue box is the field of view for zoomed-in image of panel i. (h) Model prediction of sulfur vacancies. (i) Zoomed-in STEM image of plasma-treated MoS$_2$. (j) Zoomed-in model prediction of sulfur vacancies. (k) Histogram of local intensities at S and Mo sites from as-synthesized MoS$_2$. (l) Accumulated occurrence of local intensities at model-identified vacancies and atomic sites of as-synthesized MoS$_2$. (m) Histogram of local intensities at S and Mo sites of plasma-treated MoS$_2$. (n) Accumulated occurrence of local intensities at model-identified vacancies and atomic sites of plasma-treated MoS$_2$. (o) Vacancy concentration comparison between conventional analysis (label) and FCN (model). Individual data points are obtained from images of 256×256 pixels. (p) Average vacancy concentration obtained from conventional analysis (label) and FCN (model).

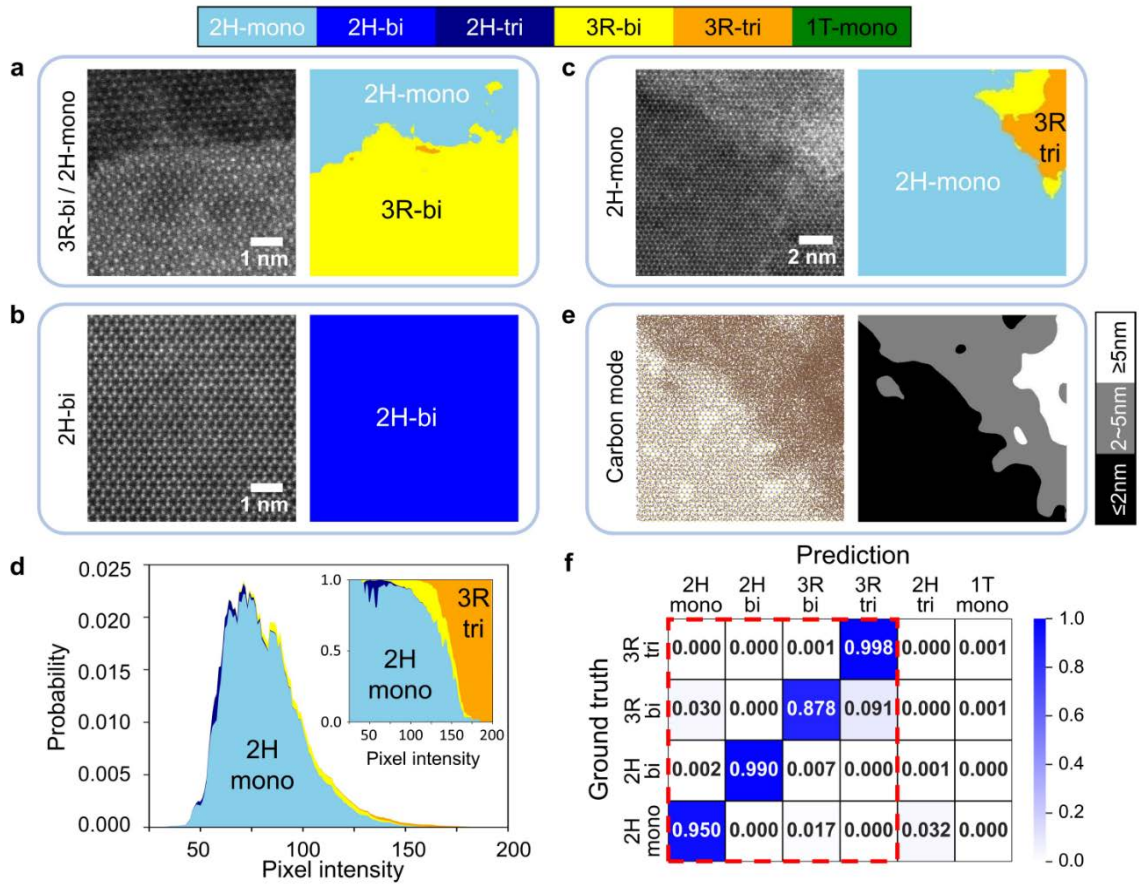

**Figure 5. Polymorph identification from experimental STEM images of MoS$_2$ using FCN.** (a) Exemplary experimental STEM images of MoS$_2$ and polymorph predictions for 2H monolayer/3R bilayer lateral junction. (b) Polymorph prediction for 2H bilayer. (c) Polymorph predictions for 2H monolayer. (d) Histogram of the local intensity of experimental 2H monolayer STEM images of MoS$_2$ with polymorph prediction distribution. The inset shows the normalized polymorph prediction as a function of local pixel intensity. (e) Estimated atomic structure of amorphous carbon (left) and carbon model prediction (right) for 2H monolayer STEM image in panel c. (f) Mean accuracy heatmap for polymorph identifications of MoS$_2$ with regions of ≤2 nm carbon contamination.

## ASSOCIATED CONTENT

**Supporting Information**.

The Supporting Information is available free of charge at https://pubs.acs.org/.

Extra notes on simulated STEM images, atom localization model, and carbon model, extra simulated STEM images, extra experimental STEM images, and extra FCN application results.

## AUTHOR INFORMATION

**Corresponding Author**

E-mail: kpkim@yonsei.ac.kr

**Author Contributions**

‡These authors contributed equally to this work.

**Notes**

The authors declare no competing interests.

## ACKNOWLEDGMENT

**Funding Sources**

We thank Prof. Hwidong Yoo for providing computing resources (YHEP server) and helpful comments and discussions. We also thank Prof. Sejung Yang for her insightful comments and advice on data pre-processing and machine learning. This work was mainly supported by the Basic Science Research Program of the National Research Foundation of Korea (NRF-

2017R1A5A1014862 and 2022R1A2C4002559) and by the Institute for Basic Science (IBS-R026-D1). Y.L. received support from the Basic Science Research Program at the National Research Foundation of Korea which was funded by the Ministry of Science and ICT (NRF-2021R1C1C2006785). Y.S.K. acknowledges support from the Priority Research Centers Program (2019R1A6A1A11053838), and the Basic Science Research Programs (2021R1A2C1004209), through the National Research Foundation of Korea (NRF), funded by the Korean government. G.H.L acknowledges the support from Creative-Pioneering Researchers Program through Seoul National University (SNU).

*12* (6), 554-561.

12. Najmaei, S.; Liu, Z.; Zhou, W.; Zou, X.; Shi, G.; Lei, S.; Yakobson, B. I.; Idrobo, J.-C.; Ajayan, P. M.; Lou, J. Vapour Phase Growth and Grain Boundary Structure of Molybdenum Disulphide Atomic Layers. *Nat. Mater.* **2013,** *12* (8), 754-759.

13. Lin, Y.-C.; Dumcenco, D. O.; Huang, Y.-S.; Suenaga, K. Atomic Mechanism of the Semiconducting-to-Metallic Phase Transition in Single-Layered $MoS_2$. *Nat. Nanotechnol.* **2014,** *9* (5), 391-396.

14. Brown, L.; Hovden, R.; Huang, P.; Wojcik, M.; Muller, D. A.; Park, J. Twinning and Twisting of Tri-and Bilayer Graphene. *Nano Lett.* **2012,** *12* (3), 1609-1615.

15. Zhao, X.; Ning, S.; Fu, W.; Pennycook, S. J.; Loh, K. P. Differentiating Polymorphs in Molybdenum Disulfide Via Electron Microscopy. *Adv. Mater.* **2018,** *30* (47), e1802397.

16. Alden, J. S.; Tsen, A. W.; Huang, P. Y.; Hovden, R.; Brown, L.; Park, J.; Muller, D. A.; McEuen, P. L. Strain Solitons and Topological Defects in Bilayer Graphene. *Proc. Natl. Acad. Sci. U.S.A.* **2013,** *110* (28), 11256-11260.

17. Yoo, H.; Engelke, R.; Carr, S.; Fang, S.; Zhang, K.; Cazeaux, P.; Sung, S. H.; Hovden, R.; Tsen, A. W.; Taniguchi, T.; Watanabe, K.; Yi, G.-C.; Kim, M.; Luskin, M.; Tadmor, E. B.; Kaxiras, E.; Kim, P. Atomic and Electronic Reconstruction at the Van Der Waals Interface in Twisted Bilayer Graphene. *Nat. Mater.* **2019,** *18* (5), 448-453.

18. Spurgeon, S. R.; Ophus, C.; Jones, L.; Petford-Long, A.; Kalinin, S. V.; Olszta, M. J.; Dunin-Borkowski, R. E.; Salmon, N.; Hattar, K.; Yang, W.-C. D.; Sharma, R.; Du, Y.; Chiaramonti, A.; Zheng, H.; Buck, E. C.; Kovarik, L.; Penn, R. L.; Li, D.; Zhang, X.; Murayama, M.; Taheri, M. L. Towards Data-Driven Next-Generation Transmission Electron Microscopy. *Nat. Mater.* **2021,** *20* (3), 274-279.

19. Dan, J.; Zhao, X.; Pennycook, S. J. A Machine Perspective of Atomic Defects in Scanning Transmission Electron Microscopy. *InfoMat* **2019,** *1* (3), 359-375.

20. Ge, M.; Su, F.; Zhao, Z.; Su, D. Deep Learning Analysis on Microscopic Imaging in Materials Science. *Mater. Today Nano* **2020,** *11*, 100087.

21. Ziatdinov, M.; Dyck, O.; Maksov, A.; Li, X.; Sang, X.; Xiao, K.; Unocic, R. R.; Vasudevan, R.; Jesse, S.; Kalinin, S. V. Deep Learning of Atomically Resolved Scanning Transmission Electron Microscopy Images: Chemical Identification and Tracking Local Transformations. *ACS Nano* **2017,** *11* (12), 12742-12752.

22. Kaufmann, K.; Zhu, C.; Rosengarten, A. S.; Maryanovsky, D.; Harrington, T.
21